\documentstyle[prc,aps,floats,epsf]{revtex}
\newcommand{\vm}[1]{\mbox{\bf#1}}
\def\thalf{{\textstyle{\frac{1}{2}}}}
\def\tquar{{\textstyle{\frac{1}{4}}}}

\def\ttwoth{{\textstyle{\frac{2}{3}}}}
\begin{document}
\preprint{NUC--MINN--98/2--T; NBI--98--40}
\title{An Effective Model for Hot Gluodynamics}
\author{G.W. Carter$^1$, O. Scavenius$^1$, I.N. Mishustin$^{1,2}$
and P.J. Ellis$^3$}
\address{$^1$The Niels Bohr Institute, Blegdamsvej 17\\
DK--2100 Copenhagen \O, Denmark\\
$^2$The Kurchatov Institute, Russian Research Center\\
Moscow 123182, Russia \\
$^3$School of Physics and Astronomy, University of Minnesota\\
Minneapolis, MN 55455, USA}
\maketitle
\begin{abstract}
We consider an effective Lagrangian containing contributions from 
glueball and gluon degrees of freedom with a 
scale-invariant coupling between the two. 
The thermodynamic potential is calculated taking into account thermal
fluctuations of both fields.  The glueball mean field dominates at low
temperature, while the high temperature phase is governed by 
low-mass gluon-like excitations. The model 
shows some similarities to the lattice results
in the pure glue sector of QCD.  In particular, it exhibits a strong first
order phase transition at a critical temperature of approximately 265
MeV when reasonable parameters are taken.
\end{abstract}

\section{Introduction}

The aim of this paper is to non-perturbatively describe the pure glue 
sector of quantum chromodynamics (QCD) with an effective Lagrangian 
containing a small number of parameters. 
In this endeavor we are guided by the requirement that the model          
reproduce the most important non-perturbative features of QCD.  At zero   
temperature the model should therefore account for the gluon condensate.  
As the temperature increases through a critical temperature, $T_c$,       
a phase transition should occur                                           
in which the condensate is melted and gluon-like 
excitations become the relevant degrees of freedom. This picture is supported 
by lattice calculations which are quite well established for the pure glue 
sector \cite{boyd96,laermann96}.  Above $T_c$ the thermodynamic ``data" has 
been fitted by several simple models ranging from one 
employing massless gluons with a low-momentum 
cut-off \cite{rischke92} to a model assuming temperature 
dependence in both the effective gluon mass and the bag constant
\cite{levai98} (see this reference for a more complete review). 

In this paper we formulate a simple effective field-theoretical model
for constituent gluons which become massive via an interaction with 
the gluon condensate. The constituent gluons are described by an Abelian
vector field, while the gluon condensate is identified with the dynamical 
glueball field which has a non-zero expectation value.
The form of the glueball potential
is well known \cite{schechter80}, fixed by the requirement that scale 
invariance be broken through generation of a trace anomaly as in QCD. 
Therefore it is surmised that the non-linearities of the actual     
QCD Lagrangian can be modeled by the glueball potential and the mass-like 
coupling term. Below we demonstrate that this simple model can indeed 
reproduce some of the features of QCD.

The plan of this paper is as follows. In Sec. II the effective Lagrangian is 
written down and the fields are decomposed into mean field and thermally
fluctuating parts. Expressions are given for the equations of motion,
masses and thermodynamic variables. Since we have to deal with a non-linear
potential we introduce a novel method of handling the
fluctuations. In Sec. III we discuss the choice of model parameters and 
present our results. Sec. IV is reserved for conclusions and outlook.

\section{The Model}
\subsection{Effective Lagrangian}

Motivated by the above considerations we write our effective Lagrangian as
\begin{equation}
{\cal L}=\thalf\partial_\mu\phi\partial^\mu\phi
-U(\phi)
-\tquar\vm{A}_{\mu\nu}\cdot\vm{A}^{\mu\nu}
+\thalf G^2\phi^2\vm{A}_\mu\cdot\vm{A}^\mu\;,\label{eq:l}
\end{equation}
where $\phi$ is the scalar glueball field
and $\vm{A}_\mu$ is the Abelian vector field with strength tensor
$\vm{A}_{\mu\nu}=\partial_\mu\vm{A}_\nu-\partial_\nu\vm{A}_\mu$. 
The dimension of the vector $\vm{A}_\mu$ is 
{\it a priori} unknown, and we define it to be 
$\nu/3$ so that the effective number of constituent gluon degrees of freedom 
is $\nu$.
The second term in Eq. (\ref{eq:l}) is the glueball potential
\begin{equation}
U(\phi)=\tquar \lambda^2\phi^4\ln\left(\frac{\phi^4}{\Lambda^4}\right)\;,
\end{equation}
where $\lambda$ is a constant and $\Lambda$, which is the only dimensionful 
parameter in the model, 
defines the vacuum glueball field, $\phi_0=\Lambda/e^{\frac{1}{4}}$.
The last term in (\ref{eq:l}) gives a masslike coupling between the 
$\phi$ and $\vm{A}_\mu$ fields with a coupling constant $G$.

Since $G$ is dimensionless, scale invariance is broken only by the
glueball potential, $U(\phi)$. This potential was constructed to reproduce
the trace anomaly of QCD in an effective theory \cite{schechter80}.
Indeed, for our Lagrangian the trace of the energy-momentum tensor  is       
\begin{equation}
\theta^\mu_\mu=4U-\phi\frac{dU}{d\phi}=-\lambda^2\phi^4\;,
\end{equation}
so that the vacuum energy          
density is $-\frac{1}{4}\lambda^2\phi_0^4\equiv-\frac{1}{4}B_0$.    
In QCD, the latter is proportional to the gluon condensate 
and is roughly known from QCD sum rules \cite{shifman79}; see 
Narison \cite{narison97} for a more recent value.
Therefore a nonzero expectation value for the glueball field 
$\langle\phi\rangle$ implies 
the presence of a gluon condensate. In the following we consider 
$\langle\phi\rangle$ to be the order parameter for 
our study of the thermodynamics and phase structure at finite temperature.
The model thus contains only four parameters: $\nu$, $B_0$, $\phi_0$ and 
$G$.
The pure glueball sector of the Lagrangian (\ref{eq:l}), which we will 
treat as a special case, 
has been discussed previously by Agasyan \cite{agasyan93}.

The Lagrangian (\ref{eq:l}) is of similar structure to the dual 
Ginzburg-Landau model \cite{mond}. The latter also assumes Abelian dominance,
for which lattice calculations offer some support, and considers three 
scalar magnetic monopole fields, rather than the single glueball field here. 
Where a quartic potential has been assumed in  Ref. \cite{mond},
we take a logarithmic potential for reasons previously explained.
In both cases the square of the constituent gluon mass is proportional     
to $\phi^2$, or equivalently the square root of the gluon condensate, 
a dependence common in other models as well \cite{perv}.                                                

\subsection{Equations of Motion and Masses}

It is convenient to take the ratio of the glueball field to its vacuum 
value, $\chi=\phi/\phi_0$. Then the  Euler-Lagrange equations of motion
are
\begin{eqnarray}
&&\phi_0^2\partial^2\chi +
2B_0\chi^3\ln\chi^2
=g^2\chi\vm{A}_\mu\cdot\vm{A}^\mu\nonumber\\
&&\partial^2\vm{A}_\mu -\partial_\mu
\left(\partial^\nu\vm{A}_\nu\right) +g^2\chi^2\vm{A}_\mu=0\;,
\label{firsteom}
\end{eqnarray}
where we have defined $g=G\phi_0$. 
The non-trivial point of our treatment is that in addition to the mean field 
we include the thermal fluctuations of the glueball and gluon fields
in a consistent way. To that end we break the glueball into mean field and 
fluctuating parts, $\chi=\bar{\chi}+ \Delta$ with 
$\langle\Delta\rangle=0$, where the angle brackets denote a thermal 
average. 
In the second of Eqs. (\ref{firsteom}) we replace $\chi^2$ 
by its thermal average 
$\langle\chi^2\rangle$ so that we can interpret the third term
as a mass term; this amounts to imposing the condition 
$\partial^\nu\vm{A}_\nu=0$, as appropriate for a vector field with three
degrees of freedom.

The field fluctuations are decomposed into plane waves and we
take the thermal average of the equations of motion, 
assuming that the thermal average of the product of glueball and gluon 
fields can be approximated by the product of their respective thermal
averages.
This gives 
the mean-field versions of Eqs. (\ref{firsteom}) for
the glueball field,
\begin{equation}
2B_0\langle \chi^3\log{\chi^2}\rangle = g^2 \bar{\chi}\langle 
\vm{A}_\mu\cdot\vm{A}^\mu\rangle\,, \label{eq:eom}
\end{equation}
and the subsequently vanishing mean gluon field, 
$\langle\vm{A}_\mu\rangle=0$. 

The dispersion relations for the gluon and glueball excitations,       
including contributions from thermal fluctuations, can be written      
\begin{equation}
e_A^2 = k^2+m_A^{*2} \quad;\quad e_{\chi}^2=k^2+m_{\chi}^{*2}\;,
\end{equation}
where the effective masses are defined to be the thermal average of 
the second derivative of the potential, {\it i.e.}
\begin{eqnarray}
\phi_0^2m_\chi^{*2}&=&-\left\langle\frac{\partial^2{\cal L}}
{\partial\Delta^2}\right\rangle=6B_0\langle\chi^2\ln\chi^2\rangle
+4B_0(\bar{\chi}^2+\langle\Delta^2\rangle)
-g^2\langle\vm{A}_\mu\cdot\vm{A}^\mu\rangle\nonumber\\
m_A^{*2}&=&\left\langle\frac{\partial^2{\cal L}}
{\partial A_\mu^{i2}}\right\rangle=g^2(\bar{\chi}^2+\langle\Delta^2\rangle)\;.
\label{eq:mass}
\end{eqnarray}
This set of equations is closed by expressing the quantities $\langle 
\vm{A}_\mu\cdot\vm{A}^\mu\rangle$ and $\langle\Delta^2\rangle$ in terms 
of the field quanta distributions.  Using standard methods one obtains 
for the vector and scalar fields, respectively,
\begin{equation}
\langle\vm{A}_\mu\cdot\vm{A}^\mu\rangle =
-\frac{\nu}{2\pi^2}\int\limits_0^\infty dk\,\frac{k^2}{e_A}n_B(e_A)
\quad\;;\quad
\langle\Delta^2\rangle = \frac{1}{2\pi^2\phi_0^2}
\int\limits_0^\infty dk\,\frac{k^2}{e_\chi}n_B(e_\chi)\;, \label{eq:sq}
\end{equation}
where  $n_B(x)=(e^{\beta x}-1)^{-1}$ is the Bose-Einstein distribution
function and $\beta=1/T$ is the inverse temperature.
The equation for the mean field
(\ref{eq:eom}) and the equations for the masses (\ref{eq:mass}) must
be solved self-consistently.

\subsection{Evaluation of the Thermal Fluctuations}

Equations (\ref{eq:eom}) and (\ref{eq:mass}) require the thermal average
of functions of $\chi$ which involve a logarithm and these are handled in the 
following manner. Consider a general function $f(\chi)$ and Taylor expand the 
fluctuations so $\langle f(\chi)\rangle=\sum_0^\infty f^{(n)}(\bar{\chi})
\langle\Delta^n\rangle/n!$, where $f^{(n)}$ denotes the $n^{\rm th}$
derivative of the function. Here we are considering the contributions from 
a single vertex, ignoring loop diagrams with two or more vertices,
as is appropriate for a mean field treatment. Simply truncating the Taylor 
expansion at low order as in Ref. \cite{agasyan93}
would be appropriate for low temperatures,
but would be inadequate for high temperatures where the mean field vanishes 
and the fluctuations are large. Fortunately we can treat 
the problem exactly. Taking the thermal average of each possible
pair of fields $\Delta$ for a given $n$-point vertex gives
$\langle\Delta^n\rangle=(n-1)!!\langle\Delta^2\rangle^{\frac{n}{2}}$ for 
$n$ even and zero for $n$ odd (see Ref. \cite{carter97} for further 
discussion). Defining a Gaussian weighting function
\begin{equation}
P(z)=\left(2\pi\langle\Delta^2\rangle\right)^{-\frac{1}{2}}
\exp-\left(\frac{z^2}{2\langle\Delta^2\rangle}\right)\;,
\end{equation}
one finds that
\begin{equation}
\langle f(\chi)\rangle=\int\limits_{-\infty}^\infty dz\,P(z)\left(
\sum_{n=0}^\infty f^{(n)}(\bar{\chi})\frac{z^n}{n!}\right)
=\int\limits_{-\infty}^\infty dz\,P(z)f(\bar{\chi}+z)\;, \label{eq:ft}
\end{equation}
where in the last step we have resummed the Taylor series. Thus the 
fluctuations enter with a Gaussian weighting and Eq. (\ref{eq:ft}) is 
straightforward to compute for any $\bar\chi$. 
By performing the series expansion one can see that
$\bar{\chi}=0$ is an
exact solution of Eq. (\ref{eq:eom}). In this case the integral 
(\ref{eq:ft}) can be 
carried out analytically \cite{gr} and Eqs. (\ref{eq:mass}) become
\begin{equation}
\phi_0^2m_\chi^{*2}=6B_0\langle\Delta^2\rangle\ln\alpha\langle\Delta^2\rangle
-g^2\langle\vm{A}_\mu\cdot\vm{A}^\mu\rangle\quad;\quad
m_A^{*2}=g^2\langle\Delta^2\rangle\;, \label{mass0}
\end{equation}
where $\ln\alpha=\frac{8}{3}-\gamma-\ln2$, with Euler's constant
denoted by $\gamma$, giving $\alpha\simeq4.0402$.

\subsection {Thermodynamics}

The grand canonical potential per unit volume can be written in a 
straightforward way:
\begin{equation}
\frac{\Omega}{V}=\thalf B_0\langle\chi^4(\ln\chi^2-\thalf)\rangle
+\tquar B_0-\thalf\phi_0^2m_\chi^{*2}\langle\Delta^2\rangle
-\frac{1}{6\pi^2}\int\limits_0^\infty dk\,k^4\!\left(
\nu\frac{n_B(e_A)}{e_A}+\frac{n_B(e_\chi)}{e_\chi}\right)\!.
\end{equation}
Here a constant second term has been
added so that $\Omega=0$ at zero temperature
and the third term subtracted so as to avoid double counting.
Notice that, apart from the constant term, all the quantities are
temperature dependent, for instance 
$\langle\chi^4(\ln\chi^2-\thalf)\rangle$ is evaluated using Eq. 
(\ref{eq:ft}) which involves the temperature-dependent quantity 
$\langle\Delta^2\rangle$ of Eq. (\ref{eq:sq}).

In order to have consistent thermodynamics $\Omega$ must be a minimum
with respect to variations in the mean field $\bar{\chi}$. Performing
the minimization we indeed find that Eq. (\ref{eq:eom})
is the necessary condition. Thus our equation of motion, our mass 
equations and the grand potential treat the thermal fluctuations in a 
coherent and well defined manner. The pressure is simply
$P=-\Omega/V$, and the energy density is easily obtained:
\begin{eqnarray}
{\cal 
E}&=&\left(1+\beta\frac{\partial}{\partial\beta}\right)\frac{\Omega}{V}=
\thalf B_0\langle\chi^4(\ln\chi^2-\thalf)\rangle
+\tquar B_0-\thalf\phi_0^2m_\chi^{*2}\langle\Delta^2\rangle\nonumber\\
&&\qquad\qquad\qquad\qquad+\frac{1}{2\pi^2}\int\limits_0^\infty dk\,k^2
\left(\nu e_An_B(e_A)+e_\chi n_B(e_\chi)\right)\,.
\end{eqnarray}
Note that
in the case $\bar{\chi}=0$, $\langle\chi^4\ln\chi^2\rangle=
3\langle\Delta^2\rangle^2 \ln\alpha\langle\Delta^2\rangle$, where $\alpha$
is the constant defined previously.

\section{Results}
\subsection{Choice of Parameters}

The model has four free parameters: $\nu$, $B_0$, $\phi_0$, and $G$.
The effective number of 
gluon degrees of freedom, $\nu$, determines the asymptotic behaviour
of the equation of state. In order to have ${\cal E}/T^4\approx 4.7$ at 
high temperature, as found on the lattice \cite{boyd96}, we need to choose 
$\nu \approx 14$ for gluons. The
standard degeneracy for massless gluons in $SU(3)$ is, of course, 16. 
We will also consider $\nu=6$ which would roughly correspond to $SU(2)$, 
as well as the case where gluons are excluded
and we have a pure glueball theory ($\nu=0$). For a given $\nu$
the critical temperature, $T_c$, of the
phase transition (see below) is largely determined by the quantity $B_0$.
For $SU(3)$ we take $\nu=14$ and
choose a value for $B_0$ of (391 MeV)$^4$ so as to reproduce the  
deconfinement temperature found in lattice calculations. 
However $B_0$ also determines the zero-temperature vacuum gluon condensate
and our value must be consistent with independent determinations of this 
quantity.
Our $B_0$ is somewhat larger than the old value of (340 MeV)$^4$ found by 
Shifman, Vainshtein and Zakharov \cite{shifman79}, but in good      
agreement with the more recent updated average of                   
$(399\pm13\ {\rm MeV})^4$ given by Narison \cite{narison97}.        
The magnitude of the vacuum energy density associated with our value of 
the gluon condensate (bag constant) $\tquar B_0\approx0.8$ GeVfm$^{-3}$. 
The third parameter, $\phi_0$, can be fixed by appealing to the vacuum 
glueball mass, $m_{\chi}^2 = 4 B_0/\phi_0^2$, which follows from 
Eq. (\ref{eq:mass}) at $T=0$.
For the glueball mass, 
values in the range 1.5 -- 1.7 GeV are suggested by data and by 
calculations \cite{pennington}; for definiteness we take 
$m_{\chi} =1.7$ GeV following Sexton {\it et al.} \cite{sexton}.
Finally in order to fix $G$ we assume that the glueball is a loosely bound 
system of two gluons
and therefore choose the effective gluon mass in vacuum, $m_A=G\phi_0=g$, 
to be $\thalf m_{\chi}$; the recent study of phenomenological gluon 
propagators \cite{lein} suggests that this is a reasonable estimate.

\subsection{Phase Transition}

Fig. 1 shows the mean glueball field, $\bar{\chi}$, for the three cases
mentioned above.  Since $\bar{\chi}$
and the other variables are essentially constant at lower temperatures
the abscissa starts at $T=100$ MeV.  A most striking 
feature of the model is that it 
exhibits a first order phase transition at a critical temperature $T_c$. 
Here the mean glueball field drops from a value of slightly less than     
unity to zero, and it remains zero for $T>T_c$. The                        
arrows on the figure indicate where the transition takes place (the      
remainder of the curves correspond to metastable or unstable branches).
The physics behind this is indicated in Fig. 2, where we plot $\Omega/V$
(relative to the value at $\bar{\chi}=0$) as a function of $\bar{\chi}$ 
for various temperatures in the vicinity of $T_c$.
Below $T_c$ there exists a local minimum of the effective potential
at $\bar{\chi}=0$, but the absolute minimum is at $\bar{\chi}\approx1$. As 
the temperature is increased the latter minimum becomes shallower 
and at $T_c$ it has the same depth
as the minimum at $\bar{\chi}=0$.  Thereafter the stable 
solution corresponds to $\bar{\chi}=0$ and ultimately the minimum at 
$\bar{\chi}\approx1$ disappears.

For the pure glueball case (dotted line in Fig. 1) the transition temperature,
$T_c=490$ MeV, is much higher than that estimated by Agasyan \cite{agasyan93}.
This is probably due to the fact that
in that work the potential was expanded to low orders, whereas here we 
have an essentially exact treatment of the thermal fluctuations.
When the glue degrees of freedom are included the transition temperature 
is much reduced and, using reasonable values for the parameters, is      
found to lie in the neighborhood of the $SU(3)$ lattice result                   
$T_c=270\pm5$ MeV for the pure glue sector of QCD \cite{laer98}.         
For our chosen parameters the first order transition for the solid       
curve occurs at $265$ MeV. For $\nu=6$ the dashed curve 
shows that the critical temperature rises to $340$ MeV. 
To compare with $SU(2)$ one should 
also take into account the scaling of the gluon condensate 
with the number of colors. Reducing the condensate by a factor of $\ttwoth$ 
results in a value of $T_c=300$ MeV. These figures are reasonable in view
of lattice calculations which yield 
a 20\% increase in $T_c$ (with considerable error) \cite{fing} when the
number of colors is changed from three to two.
However the lattice results indicate a second order transition for 
$SU(2)$ \cite{karsch},
whereas we have a first order transition, although mean field
treatments are expected to be inadequate in the neighborhood of critical     
points.

\subsection{Effective Masses}

The ratios of the effective masses to the temperature, $m_\chi^*/T$ 
and $m_A^*/T$, are displayed in 
Fig. 3 (the complicated structure of the low temperature metastable region 
for $\nu=14$ is not relevant here and for clarity is suppressed in the 
figures). The masses change little below the 
critical temperature and so the ratio drops as the temperature 
increases. Beyond the critical point the masses grow linearly with 
temperature to a good approximation. Such behavior is expected
at very high temperatures where perturbation theory is applicable.
For the solid curves ($\nu=14$) $m_A^*/T$ is quite small, 0.18, whereas 
$m_\chi^*/T$ is large, 4.5,
so that the glueball plays only a minor role beyond the transition 
temperature, as is physically expected. The reason that one inevitably 
has a large glueball mass follows from Eq. (\ref{mass0}). The gluon contribution 
to $m_\chi^*$ (second term) is positive and large at high
degeneracy $\nu$, so in order to have a low glueball mass the first term 
would have to be negative. This would be only be possible if the argument of 
the logarithm were less than 1. However for a low-mass glueball with 
$m_\chi^*/T\ll1$ the argument of the logarithm 
$\alpha\langle\Delta^2\rangle\sim[T/(1.72\phi_0)]^2$.
So in order to have a negative 
value for the logarithm for $T>T_c$, $\phi_0$ would have to 
be substantially increased compared to the chosen value of $\phi_0=180$ 
MeV. This would lead to a reduction of the vacuum 
glueball mass to an unreasonably low value. 

\subsection{Thermodynamics}

The pressure and the energy density are shown in Fig. 4,
where we plot ${\cal E}/T^4$ and $3P/T^4$. At low temperatures the energy 
density and pressure are little changed from the vacuum values.
The phase transition temperature is determined by the point at which the 
pressure curves for the two stable solutions intersect.
We note that the model predicts interesting behaviour for the pressure 
in the vicinity of this critical point.  On cooling from high temperatures the 
system can enter a supercooled metastable phase, which can even have zero 
pressure. It is conceivable that in relativistic heavy ion collisions
such behavior could allow metastable, supercooled droplets of hot gluonic 
fluid to be produced \cite{mish}.

Beyond $T_c$ the pressure and energy density in Fig. 4 increase rapidly 
to their asymptotic values.
This qualitative behavior for the pressure is in agreement with lattice 
calculations, although the approach to the asymptotic value is sharper
in our model. At high temperatures $3P\approx {\cal E}$ and the value
of these quantities for the solid and dashed curves is very close to 
that expected for an ideal gas of massless gluons with the appropriate
degeneracy, $\nu=14$ or 6, indicating that we have reached
the asymptotic regime. One aspect of this model is seemingly
in strong disagreement with the lattice data.  Namely, at the critical 
temperature there is a large latent heat and the energy density overshoots the
asymptotic value of
${\cal E}/T^4$, approaching it from above.  This is in contrast to the
lattice calculations where the latent heat is a factor of 2--3
smaller \cite{boyd96,bein} and ${\cal E}/T^4$ approaches its asymptotic 
value from below.

For $T>T_c$ the behavior can 
be understood
rather simply. The ratio of the glueball mass to temperature, $m_\chi^*/T$,
is large, while the gluon mass is small and, to a good approximation, can 
be taken to be zero.
This allows the necessary Bose integrals to be approximated easily
(see, for example, Ref. \cite{scott}). It turns out that only the 
constant term and the thermal gluon term are numerically significant for 
the pressure and energy density. Thus, one obtains expressions very similar 
to those of the bag model
\begin{equation}
\frac{3P}{T^4}\simeq\frac{\pi^2\nu}{30}-\frac{3B_0}{4T^4}\quad;\quad
\frac{{\cal 
E}}{T^4}\simeq\frac{\pi^2\nu}{30}+\frac{B_0}{4T^4}\;.\label{eq:asy}
\end{equation}
Since the pressure is approximately zero at the phase transition, the
above equation provides an estimate of the critical temperature
$T_c\approx\{45B_0/(2\pi^2\nu)\}^{1/4}$ which is accurate to 10\% or better.
The thermodynamics of the model given by Eq. (\ref{eq:asy}) can be 
contrasted with a power law fit to the lattice results \cite{boyd96}       
at high temperatures not too close to $T_c$, which gives
\begin{equation}
\left(\frac{3P}{T^4}\right)_{\rm 
latt}\simeq4.8-5.2\left(\frac{T_c}{T}\right)^{\!2.16}\quad;\quad
\left(\frac{{\cal E}}{T^4}\right)_{\rm 
latt}\simeq4.7-2.04\left(\frac{T_c}{T}\right)^{\!3.37}\;.
\end{equation}
Indeed the sign of the deviation of the energy density from the asymptotic 
value is opposite to the prediction of the bag model. This results in too 
large a latent heat, as we have mentioned earlier. We remark that the dual 
Ginsburg-Landau model appears to suffer from similar difficulties.

The qualitative features of our model are not sensitive to variation 
of the parameters within reasonable limits. For example, the primary
effect of varying the glueball-gluon coupling constant $G$ is to 
scale the gluon effective mass according to Eq. (\ref{eq:mass}).
There is little change in $T_c$, and above this temperature the 
thermodynamics is still represented by Eq. (\ref{eq:asy}) since 
$m_A^*/T$ remains small (unless $G$ is made unreasonably large). 
Similarly, the main effect of altering $\phi_0$ is to change the 
masses; increasing $\phi_0$ decreases the glueball mass and  
increases the gluon mass.  This assumes that the constant $B_0$  
has been fixed to vacuum expectations.  If we choose to alter $B_0$, 
it is the vacuum (and low-temperature) glueball mass which must change. 
This alters the critical temperature according to $T_c \sim B_0^{1/4}$,
as we have mentioned. Above $T_c$ the value of $B_0$ is inconsequential 
for the masses, while the thermodynamics still follows Eq. 
(\ref{eq:asy}).

We have also considered introducing an 
$(\vm{A}_\mu\cdot\vm{A}^\mu)^2$ term 
in the Lagrangian, which would be suggested by QCD, but this
does not appear to improve the situation. However it is worth noting that 
the simple replacement $m_A^{*2}\rightarrow m_A^{*2}+m_0^2$, where $m_0$ is
a constant mass of order 500 MeV, can alter the predictions. With some 
modest adjustment of the other parameters, the energy density and pressure 
above $T_c$ can be brought into semi-quantitative agreement with the 
lattice data. In particular, the energy density no longer overshoots the 
asymptotic value. 
This modification, however, is inconsistent with QCD in that 
the corresponding term in the Lagrangian, 
$\thalf m_0^2\vm{A}_\mu\cdot\vm{A}^\mu$, introduces a dimensional parameter 
and consequentially gives an unwanted contribution to the trace anomaly. 
Thus such an addition contradicts our initial approach. 
It does however strongly suggest that the problem with
our model is that above $T_c$ the masses immediately become proportional to the 
temperature and some important physics has not been accounted for in our 
description of the masses in the region $T_c <T < 2T_c$. 

\section{Conclusions}

In conclusion, we have examined a very simple Lagrangian model for the
pure glue sector of QCD and determined its thermal properties
using a novel treatment of the thermal fluctuations.  In accord with 
physical expectations, the model shows a phase transition between a 
low-temperature, glueball-dominated regime and a high-temperature phase 
dominated by low-mass gluon-like excitations. Sensible           
parameters yield a transition temperature in agreement with 
lattice simulations. Beyond $T_c$ the qualitative behavior of the pressure
is reasonable, but the energy density in the neighborhood of the phase 
transition disagrees with the lattice data. 
This pathology appears to be a result of the the low effective             
gluon mass immediately after the phase transition.                         
So while the thermodynamics are reasonable at low and high                 
temperatures, the lattice results reveal our model to be incomplete        
just above the critical temperature.                                       
Nevertheless, we conclude that our simple Lagrangian indeed captures
important non-perturbative features of QCD.

In the future it would be interesting to 
consider modifications suggested by the color dielectric model which 
simulates confinement. It would also be worthwhile to
consider the inclusion of quark degrees of freedom. Such effective 
models may be particularly useful for simulations of time-dependent 
processes and the effects of finite baryon density which are yet intractable 
on the lattice.

\section*{Acknowledgements}

We thank \'Agnes M\'ocsy for assistance with some of the calculations.
This work was supported in part by the US Department of Energy under grant 
number DE-FG02-87ER40328, by the US National Science Foundation under travel 
grant INT-9602108 and by the Carlsberg Foundation, Denmark.

\newpage

\begin{figure}[t]
\setlength\epsfxsize{10cm}
\centerline{\epsfbox{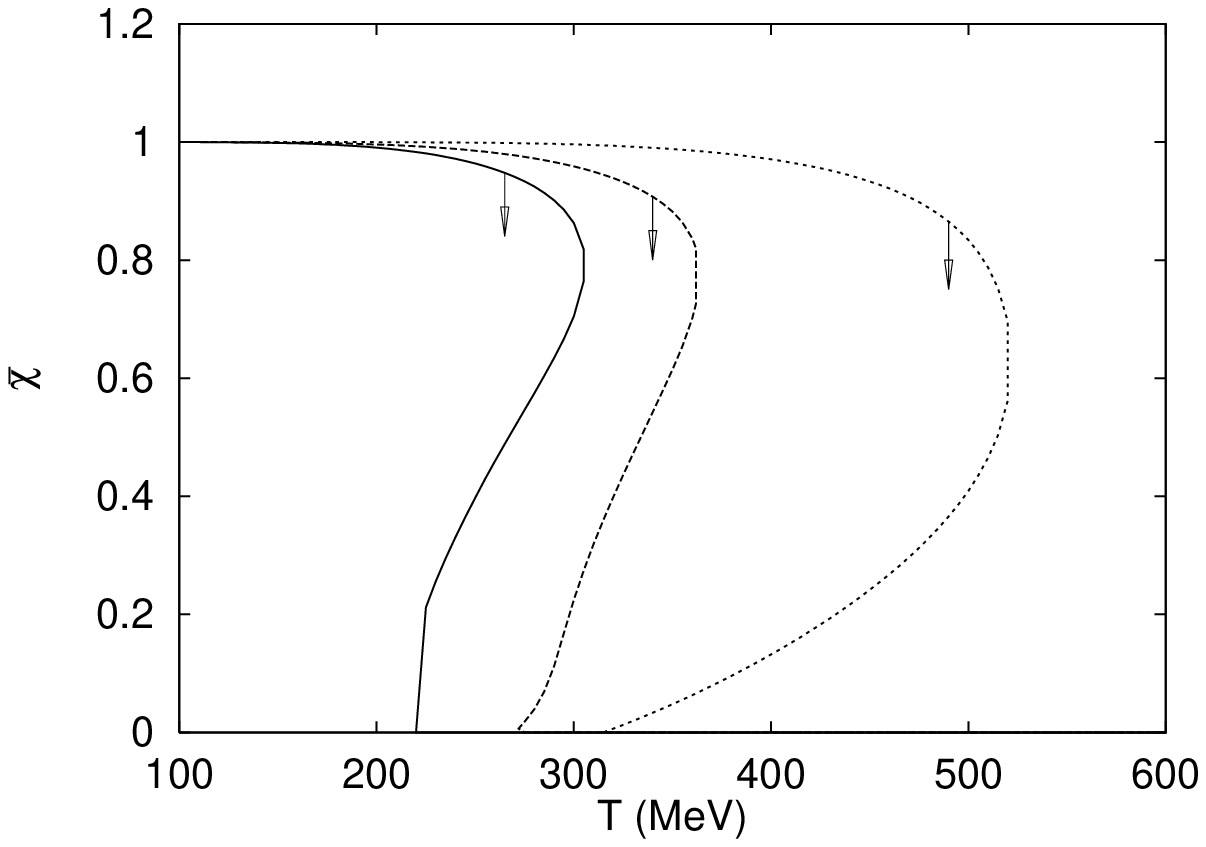}}
\caption{ 
The mean glueball field, $\bar{\chi}$, as a function
of temperature, $T$. For the dotted curve the gluon  
field is neglected, while the dashed and solid curves correspond
to degeneracies, $\nu$, of 6 and 14 for the gluon field, respectively.
The arrows indicate where a phase transition takes place and the 
thermodynamically stable phase becomes $\bar{\chi}=0$.
}
\end{figure}
\begin{figure}[b]
\setlength\epsfxsize{10cm}
\centerline{\epsfbox{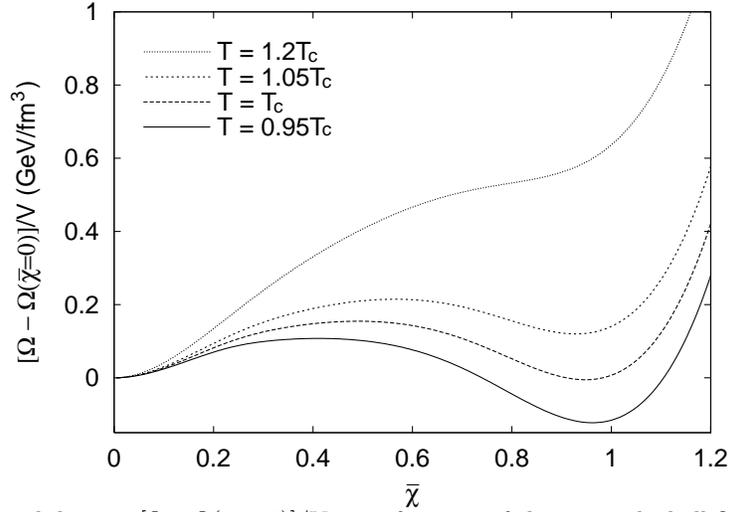}}
\caption{ 
The grand potential density, $[\Omega-\Omega(\bar{\chi}=0)]/V$,
as a function of the mean glueball field, $\bar{\chi}$, for temperatures,
$T$, in the vicinity of the critical temperature $T_c$.}
\end{figure}

\begin{figure}[tb]
\begin{center}
\leavevmode
\setlength\epsfxsize{10cm}
\epsfbox{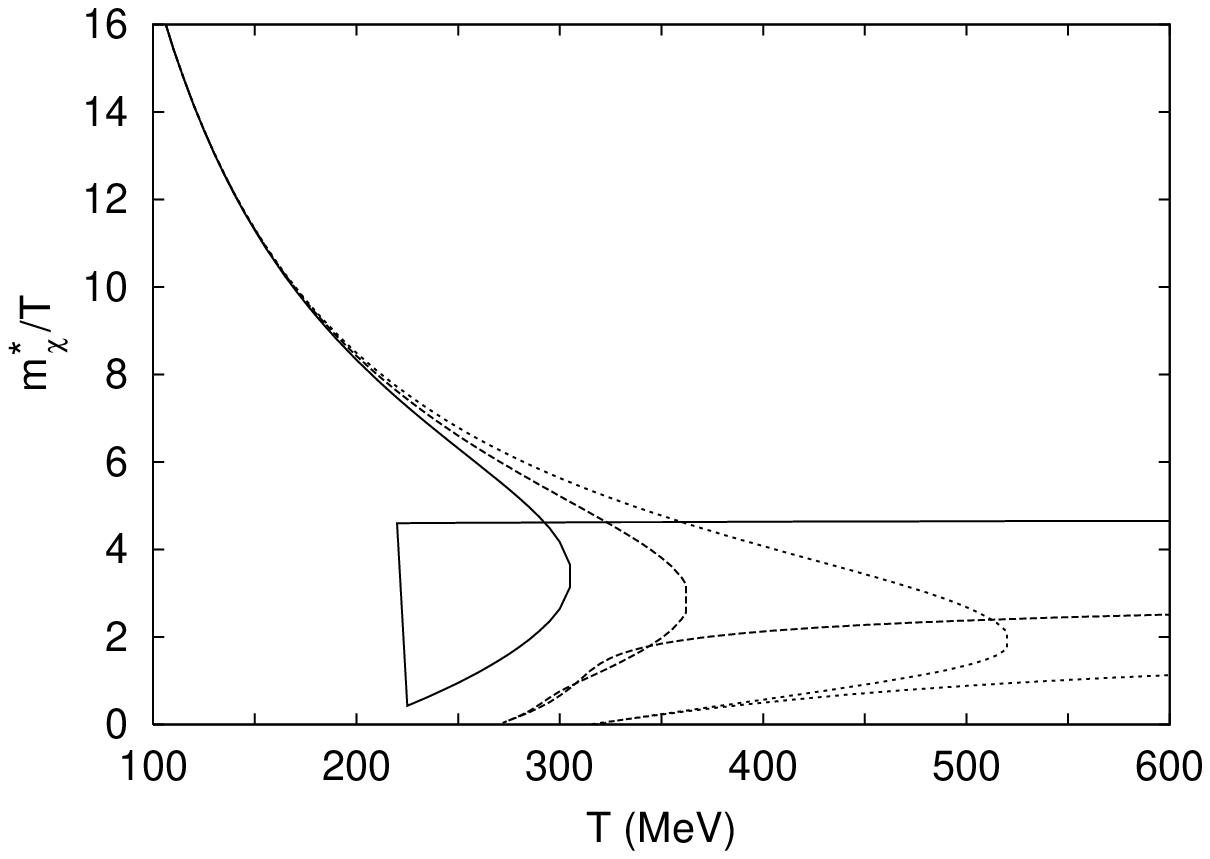}
\hspace{1cm}
\setlength\epsfxsize{10cm}
\epsfbox{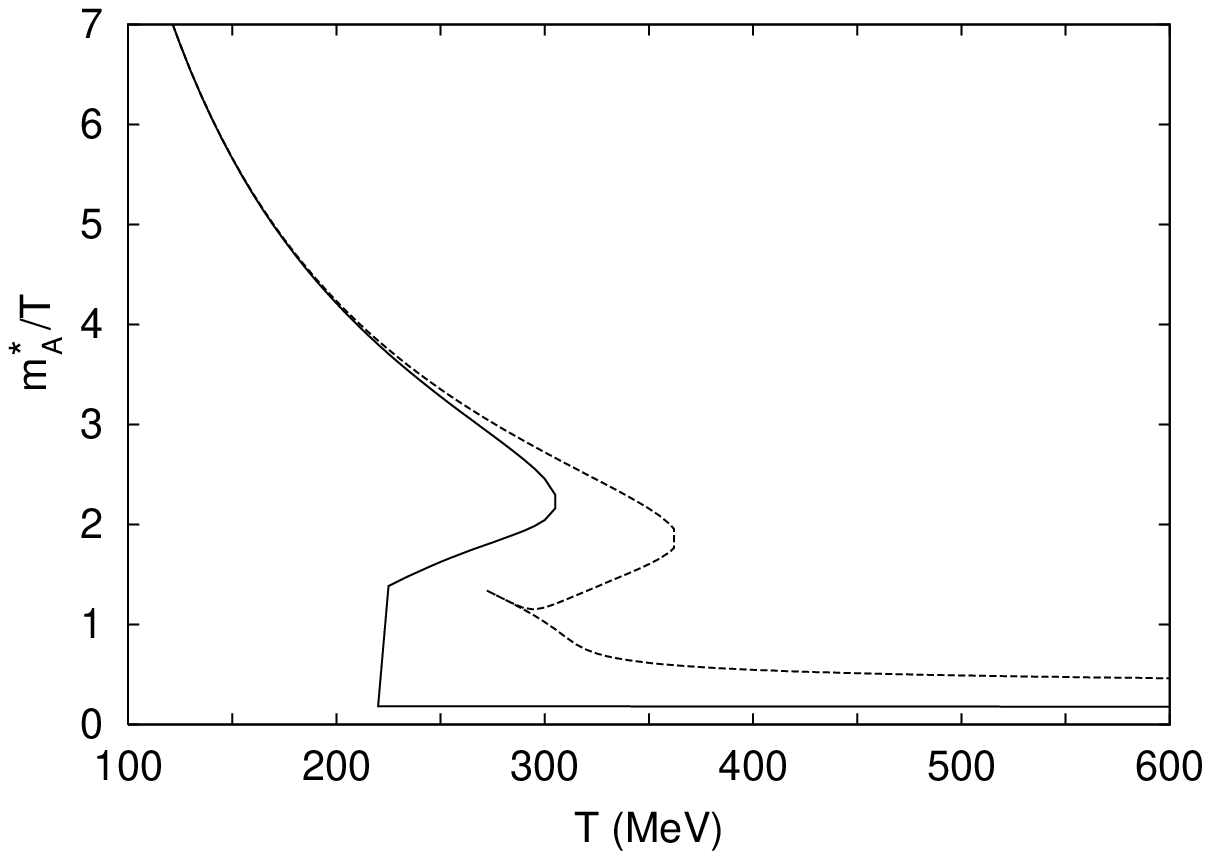}
\end{center}
\caption{
The effective masses of the glueball and gluon in 
units of the temperature, $m_\chi^*/T$ and $m_A^*/T$, as a function of
temperature. See caption to Fig. 1 for the meaning of the curves.
}
\end{figure}

\begin{figure}[tb]
\begin{center}
\leavevmode
\setlength\epsfxsize{10cm}
\epsfbox{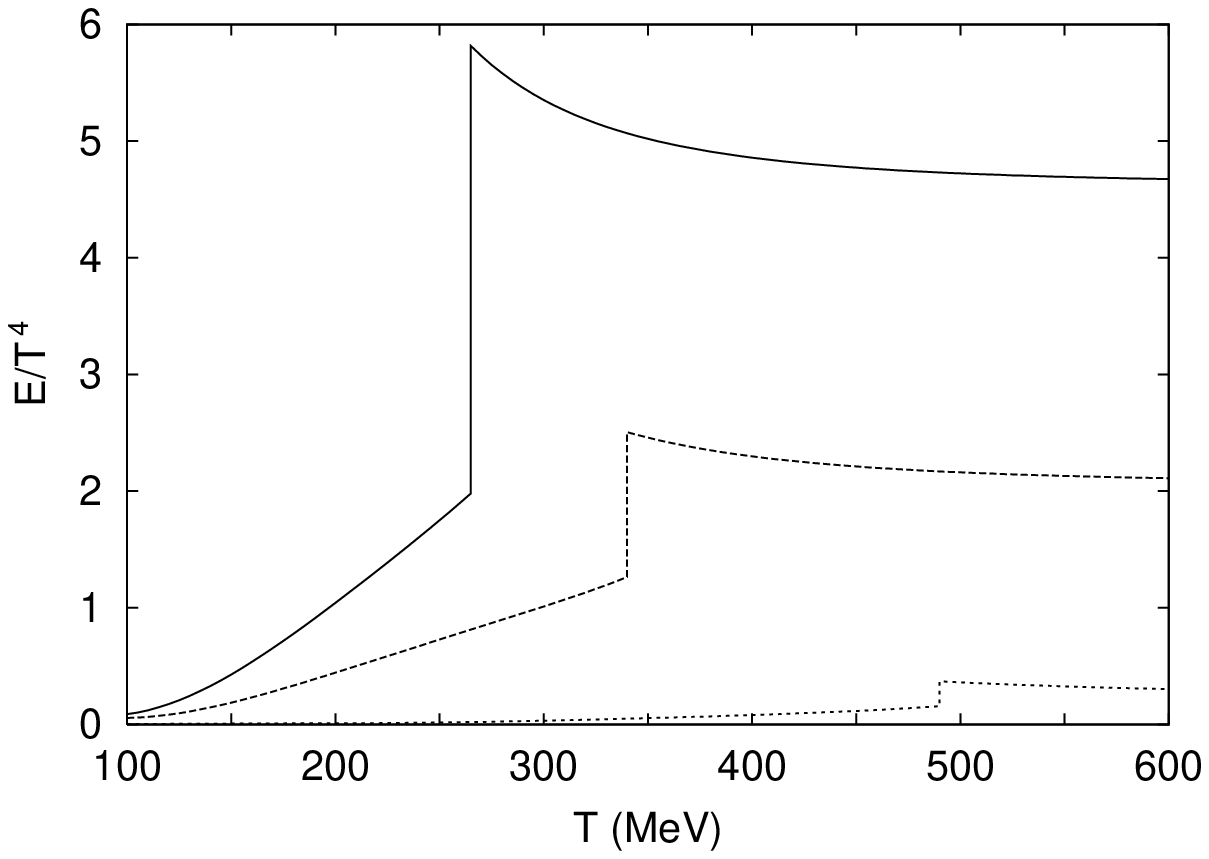}
\hspace{1cm}
\setlength\epsfxsize{10cm}
\epsfbox{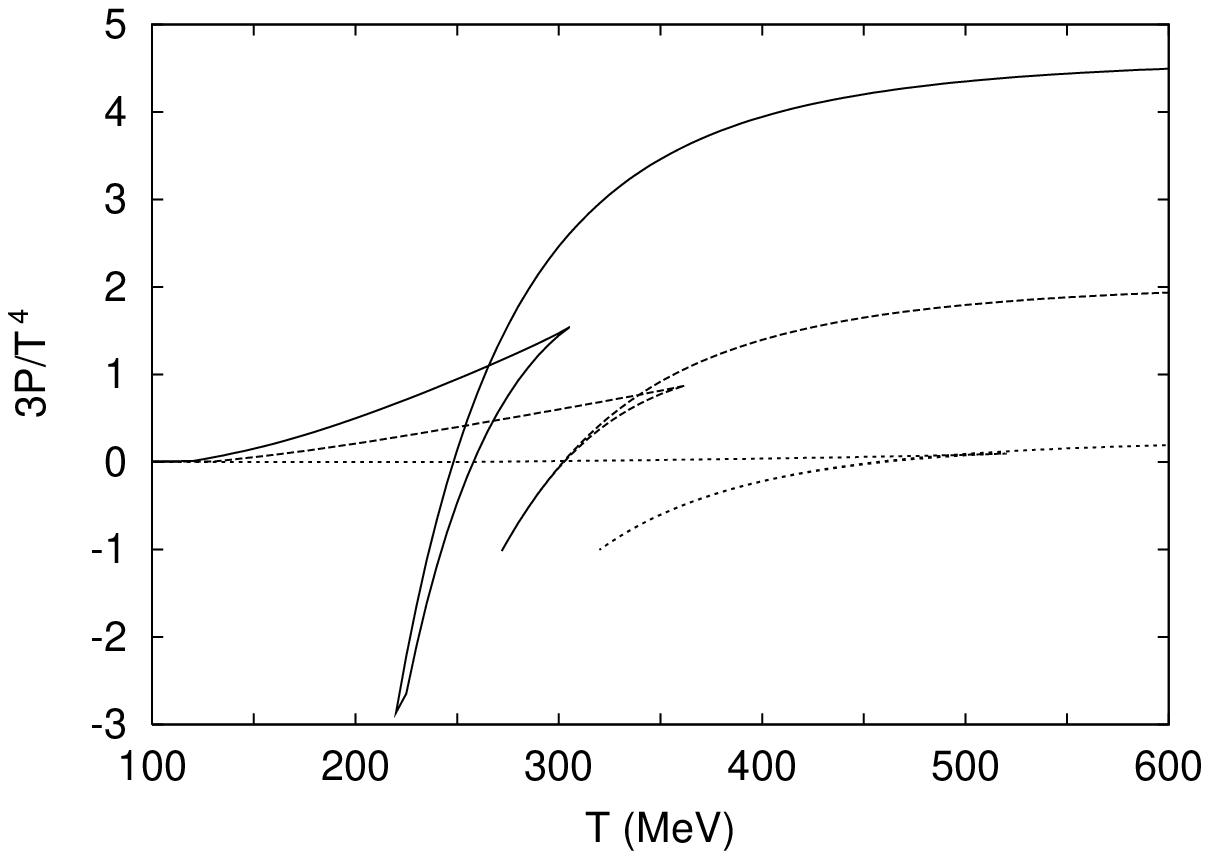}
\end{center}
\caption{
The thermodynamic quantities 
${\cal E}/T^4$ and $3P/T^4$ as a function of temperature. 
See caption to Fig. 1 for the meaning of the curves.
}
\end{figure}

\end{document}